\documentclass[11pt,a4paper]{article}

\usepackage{jcappub}
\usepackage{color}
\usepackage[T1]{fontenc} 
\usepackage{lmodern}
\usepackage{url}
\usepackage{graphicx}
\usepackage[caption=false]{subfig}
\usepackage{amsmath,amssymb}
\usepackage{amsfonts}
\usepackage{rotating}
\usepackage{epstopdf}
\usepackage{epsfig}
\usepackage{latexsym}
\usepackage{bm}
\usepackage[utf8]{inputenc}
\usepackage{float}
\usepackage{ulem}
\usepackage{hyperref}
\hypersetup{ setpagesize=false,  colorlinks=true, linkcolor=blue, citecolor=red, linktocpage=true, urlcolor=blue, hypertexnames=true}
\usepackage{natbib}

\newcommand{\mpl}{M_{\text{pl}}}

\newcommand{\dphiv}{\dot{\phi}_{0}}

\def\<{\langle\,}
\def\>{\,\rangle}
\newcommand{\paren}[1]{\left( #1 \right)}

\newcommand{\Fig}[1]{figure~\ref{#1}}
\newcommand{\Eq}[1]{Eq.~(\ref{#1})}
\newcommand{\Eqs}[2]{eqs.~(\ref{#1}) and (\ref{#2})}
\newcommand{\Sec}[1]{section~\ref{#1}}

\newcommand{\hinf}{H_{\rm inf}}
\newcommand{\Omgw}{\Omega_{\rm GW}^{(0)} h^2}
\newcommand{\dOmgw}{\delta\Omega_{\rm GW}^{(0)} h^2}

\newcommand{\rhox}[1]{\rho_{\rm #1}}
\newcommand{\f}[1]{f_{\rm #1}}

\graphicspath{{figures/}}

\title{ {\huge ~~~~~~ Large Primordial Fluctuations in \\ Gravitational Waves from Phase Transitions} }
\author{\Large Arushi Bodas}
\author{and Raman Sundrum}
\affiliation{Maryland Center for Fundamental Physics, Department of Physics,\\University of Maryland, College Park, MD 20742, USA}
\emailAdd{arushib@terpmail.umd.edu}
\emailAdd{raman@umd.edu}

\abstract{It is well-known that first order phase transitions in the early universe can be a powerful source of observable stochastic gravitational wave backgrounds. Any such gravitational wave background must exhibit large-scale anisotropies
 {\it at least} as large as those seen in the CMB $\sim 10^{-5}$, providing a valuable new window onto the (inflationary) origins of primordial fluctuations. While significantly larger fractional anisotropies are possible (for example, in multi-field inflation) and would be easier to interpret, it has been argued that these can only be consistent with CMB bounds if the gravitational wave signal is correspondingly smaller. In this paper, we show that this argument, which relies on assuming  radiation dominance of the very early universe, can be evaded if there is an era of early matter dominance of a certain robust type. This allows large gravitational wave anisotropies to be consistent with observable signals at proposed future gravitational wave detectors. Constraints from the CMB on large scales, as well as primordial black hole and mini-cluster formation on small scales, and secondary scalar-induced gravitational waves are all taken into account.
}

\begin{document}
\maketitle
\section{Introduction}

Well-motivated particle physics beyond the Standard Model (BSM) can readily undergo first-order phase transitions in the early universe, which can be  powerful sources of observable stochastic gravitational wave backgrounds (GWB)  (see \cite{caprini2016science,Mazumdar:2018dfl} for a review). For example, a critical temperature in the (multi-)TeV range would result in GWB in the frequency range of the LISA detector \cite{amaro2017laser}. 
The frequency spectrum of a GWB would encode valuable information about the BSM dynamics. Complementary to the frequency spectrum, it was
shown in \cite{Geller:2018mwu} that such a GWB  would necessarily exhibit large-scale anisotropies, analogous to those of the CMB.
GWB anisotropies in other context were discussed in \cite{Olmez:2011cg,Bethke:2013aba,Bethke:2013vca,Jenkins:2018nty,Bartolo:2019oiq,Bartolo:2019yeu,ValbusaDallArmi:2020ifo}.
In the inflationary paradigm these would reflect quantum fluctuations in inflation-era fields, giving us an invaluable new window into their poorly understood dynamics.
In particular, if there are multiple light fields during inflation, Ref.~\cite{Geller:2018mwu} also showed that the GWB anisotropies could have a significant isocurvature component, very different from the standard adiabatic perturbations of the CMB and Large Scale Structure (LSS).
Refs.~\cite{Kumar:2021ffi,Bodas:2022zca} have explored the potential of such isocurvature GWB maps from phase transitions to  probe  early universe physics.

The presence of isocurvature in GWB would imply the existence of another light quantum field during inflation.
However, to be readily distinguished and interpreted the isocurvature would have to be larger than the CMB anisotropy $\sim  10^{-5}$.
This is because the GWB would receive an irreducible contribution from the adiabatic perturbations  through the gravitational Sachs-Wolfe effect \cite{Contaldi:2016koz,Bartolo:2019oiq,Kumar:2021ffi,sachs1967perturbations}.
Therefore, large fractional anisotropies in the GWB $> 10^{-5}$ would be ideal for revealing new inflationary physics. 

For example, a well-motivated candidate for a light spectator field during inflation is an (unstable) axion-like particle (ALP) with an initial misalignment from its minimum (see \cite{ringwald2014axions,marsh2016axion} for a review of ALPS).
The overall level of anisotropy is then given by  the ratio of the inflationary Hubble scale to this field-misalignment.
If the Hubble constant during inflation can be inferred independently, through the detection of primordial tensor modes for example \cite{baumann2009tasi}, then the measurement of GWB isocurvature would give us the initial misalignment. Indeed, in high-scale inflation where the Hubble constant is taken to be $\sim 10^{-5}$ times the Planck scale \cite{Akrami:2018odb}, the ALP fluctuations are likely to be 
$> 10^{-5}$ if the ALP misalignment is constrained to being sub-Planckian.


However, there is a significant challenge for large GWB anisotropies to even be detectable. A distinct isocurvature GWB requires that there be two mostly decoupled sectors at the time of the phase transition and GWB production. One sector with adiabatic perturbation is reheated by the inflaton, and another undergoing the phase transition is reheated by a separate light field, in our case an ALP. The phase-transitioning sector in such a case must be significantly subdominant at recombination to avoid its impact on the CMB, which is highly constrained from observations. 
In the case of standard radiation dominance in the early universe in both sectors, this subdominance implies a strong suppression of the GWB signal \cite{Geller:2018mwu},
making detection of large anisotropies extremely difficult at upcoming or proposed detectors.

In this work, we will show that the trade-off between large fractional anisotropy and detectability of GW signals can be evaded if there is a period of early matter dominance (eMD)\footnote{Early here means before Big Bang Nucleosynthesis.} in the 
cosmological history. In this scenario, the phase-transitioning sector can be dominant at the time of the phase transition, avoiding suppression in GW signal at its production time. 
A later stage of eMD in the adiabatic sector will dilute the phase-transitioning sector such that its contribution to CMB anisotropies is subdominant. It would necessarily also dilute the overall GWB, but would leave the anisotropic component unsuppressed in absolute size, plausibly within the sensitivity of upcoming detectors. 

The paper is organized as follows. In section~\ref{sec:ghostmodel}, we review the original model of isocurvature GWB from a phase transition from \cite{Geller:2018mwu}, in which the phase transition and adiabatic sectors of the 
very early universe are always radiation-dominated,
prior to merging into a single sector via (late) particle decays. We
re-examine why, with radiation dominance, consistency with CMB constraints predicts that large fractional anisotropies in the GWB must come at the expense of suppressed signal strength, both in the isotropic and anisotropic components. In section~\ref{sec:emdmodel}, we introduce our new eMD model and show how a large fractional anisotropy in the GWB is consistent with CMB constraints, along the lines sketched above. We also show that the eMD dilution of the GWB results in only a modest redshift of the GW frequency spectrum relative to the original model. In section~\ref{sec:gwcrosstalk}, we consider the gravitational interactions between the phase transition and adiabatic sectors prior to their merging to see that they do no affect our conclusions in section~\ref{sec:emdmodel} on large scales, but also to consider new constraints on small scales from primordial black hole production, small-scale structure formation, and scalar-induced gravitational wave production at second order.  In section~\ref{sec:benchmarks}, we present plausible benchmarks in our model with different fractional GWB anisotropies and the associated isotropic and anisotropic signal strengths. We compare with the analogous benchmarks in the original radiation-dominated model to illustrate the significant gains in signal strength, in both the isotropic and anisotropic components. We conclude in section~\ref{sec:discussion} with a short discussion of the significance of our results.

\section{The radiation-dominated model of isocurvature GWB} \label{sec:ghostmodel}
\begin{figure}[h]
	\centering		\includegraphics[width=0.7\linewidth,trim={5cm 4cm 5cm 5cm},clip]{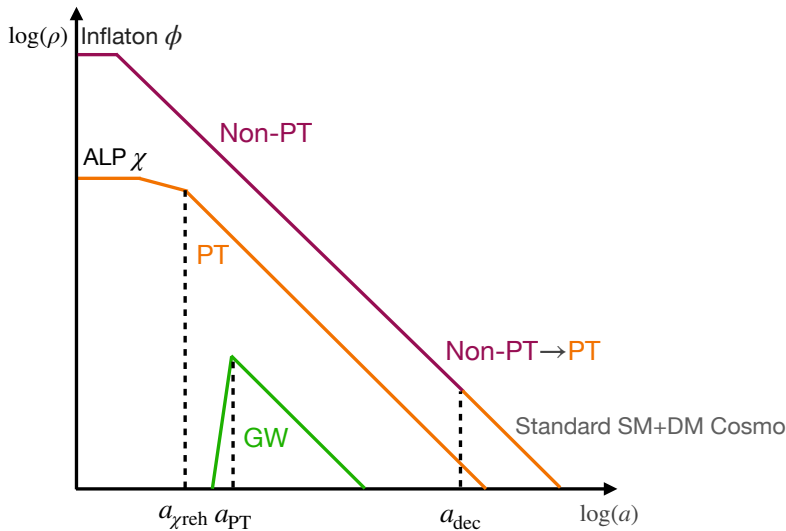}
	\caption{A schematic of the model discussed in  \Sec{sec:ghostmodel} showing energy densities in various sectors as a function of the scale factor (time).
    The decay products of both the inflaton and  ALP $\chi$ are always radiation dominated in this model.} 
	\label{fig:ghost}
\end{figure}
In this section, we review the model proposed in ref~\cite{Geller:2018mwu} that produces isocurvature GWB.
Consider an additional light scalar field $\chi$ during inflation. 
It is taken to be initially misaligned from its minimum (taken to be at $\chi = 0$) to $\chi_0 \gg \hinf$, where $\hinf$ denotes the inflationary Hubble constant.
Such a misalignment is robust as the field remains stuck due to the inflationary Hubble friction if its mass $m_{\chi} < \hinf$.
A well-motivated example of such a light field is an axion-like particle (ALP),  which is a generic prediction of many BSM theories (see \cite{ringwald2014axions,marsh2016axion}).
During inflation, the energy density in $\chi$, $V(\chi_0) \sim m_{\chi}^{2} \chi_{0}^{2}$ is subdominant to that of the inflaton $\phi$, which primarily drives inflation. 
After inflation, when the dropping Hubble constant matches $m_{\chi}$,
the $\chi$ field starts oscillating around its minimum and reheats some extension of the standard model (SM), including dark matter (DM), at $t_{\chi{\rm reh}}$. We will assume that this extension of the SM 
undergoes a first-order phase transition (PT)  in the multi-TeV range, capable of producing an observable GWB, as is known to take place in many explicit SM extensions (see \cite{caprini2016science,Mazumdar:2018dfl}). 
If the reheat temperature from $\chi$ decay is higher than the PT temperature, the decay products of $\chi$ will undergo this phase transition. 
We will therefore refer to this sector emerging from $\chi$ decay as the ``PT sector''.


The inflaton on the other hand is taken to reheat another sector that interacts very weakly with the PT sector and does not participate in the phase transition, henceforth referred to as the ``non-PT sector''.
The non-PT sector is
taken to decay to the SM $+$ DM at $t_{\rm dec}$ sometime after the phase transition, but before DM decoupling. 
In this way, after the decay there is a single SM$+$DM sector which dominantly originates from inflaton $\phi$ reheating, assuming that the PT sector is subdominant to the non-PT sector as depicted in \Fig{fig:ghost}. On the other hand, 
the GWB dominantly originates from $\chi$ reheating. Therefore it is possible for the fluctuations in GWB to be distinct from those in the SM$+$DM, as we detail below. 


\subsection{Fluctuations}
Here, we relate the anisotropies in the CMB and GWB to the independent primordial fluctuations $\zeta_{\phi}$ and $\zeta_{\chi}$\footnote{A more relevant quantity to measure isocurvature is ${\cal S}_{\chi} =  \zeta_{\chi} - \zeta_{\phi}$. However, for $\zeta_{\chi} \gg \zeta_{\phi}$, which is of interest to us in this work, we can simply take ${\cal S}_{\chi} \approx \zeta_{\chi}$.},
\begin{align}
    \zeta_{\phi}(k) =  \frac{H_{\rm inf} \delta \phi(k)}{\dphiv} , \quad  \zeta_{\chi}(k)  = \frac{2 \delta \chi (k)}{3 \chi_{0}} \,,
\end{align}
where $\delta \phi \sim \delta \chi \sim H_{\rm inf}$, and the expressions are evaluated at the horizon exit of the co-moving mode $k$.
The fluctuations in GWB will be dominated by the PT sector (reheated by $\chi$), $\delta_{\rm GW} \sim  \delta_{\rm PT} \sim \zeta_{\chi}$. 
More precisely, the anisotropies can be written in terms of the gauge-invariant fluctuations as \cite{Kumar:2021ffi, Malik:2008im}
\begin{align}\label{eq:delgwghost}
    \delta_{\rm GW} = 4 \zeta_{\chi} \left(1-\frac{4}{3}f_{\rm PT}\right) -\frac{16}{3}\zeta_{\phi}  \,,
\end{align}
where 
\begin{align}
    \f{PT} \equiv \frac{\rho_{\rm PT}}{\rho_{\rm total}}
\end{align}
is the fraction of energy density in the PT sector compared to the total energy density.
Note that $\zeta_{\phi}\sim 10^{-5}$ sets the minimum $\delta_{\rm GW}$.
This comes from the Sachs-Wolfe (SW) effect, which is dominated by the fluctuations of the non-PT sector. 
A small $\zeta_{\chi} < 10^{-5}$ would be swamped by the SW contribution, leaving the GWB to be predominantly adiabatic.
On the other hand, large isocurvature in GWB coming from $\zeta_{\chi} > 10^{-5}$,  would clearly imply the existence of a new light quantum field during inflation.
Hence, we will focus on the range $ \zeta_{\chi} > 10^{-5}$.

The fluctuations in the SM plasma (and hence in the CMB and matter distribution) after $t_{\rm dec}$
are a weighted superposition of the fluctuations in the PT sector and those in the decay
products of the non-PT sector, $\delta_{\gamma}  \sim  \delta_{\rm nPT}(t_{\rm dec})  + f_{\rm PT} \delta_{\rm PT} $.
In terms of the primordial fluctuations, 
\begin{align}\label{eq:delcmbghost}
    \delta_{\gamma} \approx  -\frac{4}{5} \zeta_{\phi} -\frac{4}{5}f_{\rm PT} \zeta_{\chi}.
\end{align}
As expected, the contribution of $\zeta_{\chi}$ to CMB can be kept small by choosing sufficiently small $\f{PT}$. 
From the CMB observations \cite{Planck:2018vyg}, $\delta_{\gamma} \simeq 3.6 \times 10^{-5}$, giving a constraint 
\begin{align}\label{eq:fptconstraint}
    \f{PT} \, \zeta_{\chi} \leq 4.5 \times 10^{-5} \,.
\end{align}
Therefore, since we are interested in $\zeta_{\chi} \gg 10^{-5}$, the PT sector must be sufficiently subdominant, $f_{\rm PT} \ll 1$.
This leads to the simplification of \Eq{eq:delgwghost}, 
\begin{equation}\label{eq:delgwghostsimplified}
    \delta_{\rm GW} \approx 4 \zeta_{\chi}.
\end{equation}
Additionally, if $\zeta_{\chi}$ contains qualitative features which are not observed in CMB, the constraint on $\f{PT} \,\zeta_{\chi}$ will be even stronger, requiring yet smaller $f_{\rm PT}$.

\subsection{Strength of GWB and its anisotropies}\label{sec:ghostgwstrength}
It is known that the GW production is suppressed when the PT sector constitutes a smaller fraction $f_{\rm PT}$ of the total energy density at the time of the phase transition.
In terms of the fraction of energy in gravitational waves compared to the critical energy density at the time of production $t_{\rm PT}$,
the suppression goes as 
\begin{align}\label{eq:gwPTghost}
    \Omega_{\rm GW}(t_{\rm PT}) \equiv \frac{\rhox{GW}(t_{\rm PT})}{\rhox{total}(t_{\rm PT})} \propto  \f{PT}^{2} 
\end{align}
for dominant sources of gravitational wave production, such as bubble collisions and acoustic waves \cite{Breitbach:2018ddu,Fairbairn:2019xog}.
The proportionality constant is determined by the microphysics of the phase transition.
Heuristically, Einstein equation for the gravitational radiation takes the form $\omega_{*}^{2} h_{\rm GW} \sim \rho_{\rm PT}/\mpl^2$, where $\omega_{*}$ is the typical GW frequency at production, $h_{\rm GW}$ is the amplitude of the metric perturbation corresponding to the gravitational wave, and $\mpl^{2} = (8 \pi G)^{-1}$ is the reduced Planck mass.
Since the energy density in GW goes as  $\rho_{\rm GW} \sim \omega_{*}^{2} \mpl^{2} h_{\rm GW}^{2}$, it follows that $\Omega_{\rm GW}(t_{\rm PT})\sim f_{\rm PT}^2 H^2/\omega_{*}^{2} $. Time dependence is set by the $H$ scale so that $\omega_{*}/H$ is determined by the microphysics, independent of cosmological abundances\footnote{There are milder dependencies on $\f{PT}$ which we ignore since we are tracking leading order effects.}, leading to \Eq{eq:gwPTghost}. 

We specialize to the plausible and relatively tractable scenario in which the gravitational waves are produced dominantly from bubble wall collisions in the thin wall regime\footnote{In the case of significant coupling between the PT plasma and bubble walls, the majority of latent heat can be transferred to the plasma making sound waves the dominant source of gravitational waves. However, it turns out that for the parameters that we choose as benchmark, i.e. $\alpha_{\rm PT} \sim 1 $ and $\tilde{\beta} \sim 10$, the sound wave contribution will be similar in strength to the estimate in \Eq{eq:ptgwestimateGhost} even if all the latent heat is transferred to the kinetic energy of the plasma \cite{Hindmarsh:2017gnf}.}.
The energy density in the gravitational waves dilutes like radiation after production. 
Then the GWB spectrum today is given by \cite{caprini2016science}
\begin{equation}\label{eq:gwbubbles}
    \begin{aligned}
    \Omega_{\rm GW}^{(0){\rm bub}} h^2  & \approx 1.67 \times 10^{-5} \tilde{\beta}^{-2} \f{PT}^{2} \,
     \left(\frac{\kappa \alpha_{\rm PT}}{1+ \f{PT} \alpha_{\rm PT}}\right)^2 
    \left(\frac{100}{g_{*}(t_{\rm PT})}\right)^{1/3} \left(\frac{0.11 v_{w}^{3}}{0.42+v_{w}^{2}}\right) S_{\rm bub}(\omega),  
\end{aligned}
\end{equation}
where the frequency dependence is 
\begin{align}
    S_{\rm bub}(\omega) &= \frac{3.8 (\omega/\omega_{\rm peak})^{2.8}}{1+2.8(\omega/\omega_{\rm peak})^{3.8}} \,.
\end{align}
Here, $\alpha_{\rm PT} $ is the ratio of the latent heat released compared to the energy density of the PT plasma before the phase transition, and $\kappa$ is the efficiency of the latent heat transfer to bubble walls. We will take $\alpha \sim 1$, such that the efficiency $\kappa \sim 1$, and the bubble wall velocity $v_w \approx 1$. 
The peak frequency today $\omega_{\rm peak}$ can be estimated by redshifting the typical frequency at the time of phase transition $\omega_*$,
given by $  \omega_{*} \approx 0.23 \, H(t_{\rm PT}) \,\tilde{\beta} $.
We choose $\tilde{\beta} =10$ for all our benchmarks. 
$g_{*}$ corresponds to the effective number of relativistic degrees of freedom, and we take $g_{*}(t_{\rm PT})\sim 100$~\footnote{In reality, $g_{*}(t_{\rm PT}) > 100$ depending on the exact extension of the SM undergoing the phase transition. However, this is a small effect considering the $g_{*}^{-1/3}$ dependence in \Eq{eq:gwbubbles}.}.
With this choice of parameters, the expression in \Eq{eq:gwbubbles} at the peak frequency is simplified to
\begin{equation}\label{eq:ptgwestimateGhost}
    \begin{aligned}
    \Omega_{\rm GW}^{(0)} h^2 
     \approx & \, 1.3 \times 10^{-8}  \f{PT}^{2} \,,
\end{aligned}
\end{equation}
where we have taken $\f{PT} \ll 1$. 


Following \Eq{eq:delgwghostsimplified}, the  size of the fluctuations in the GWB, which is relevant for detection, is 
\begin{align}
    \dOmgw \approx 
    \Omgw \times 4 \zeta_{\chi} \approx 5 \times 10^{-8}   \f{PT}^{2} \zeta_{\chi}\,.
\end{align}
The constraint of \Eq{eq:fptconstraint} gives us an upper bound on the inhomogeneities of the GWB, 
\begin{align}\label{eq:delomegagwPTghost}
    \dOmgw  \leq 
    2.2 \times 10^{-12}   \f{PT} \,. 
\end{align}
Note that despite relative anisotropy $\delta \rho_{\rm GW}/ \rho_{\rm GW} \sim \zeta_{\chi}$ being large $> 10^{-5}$, the absolute  signal strength of the fluctuations is suppressed by one power of $f_{\rm PT}$, making it more difficult to detect.

\section{Isocurvature GWB with an era of early matter dominance (eMD)}\label{sec:emdmodel}
\begin{figure}[h]
	\centering
		\includegraphics[width=0.7\columnwidth,trim={5cm 4.5cm 4cm 5.5cm},clip]{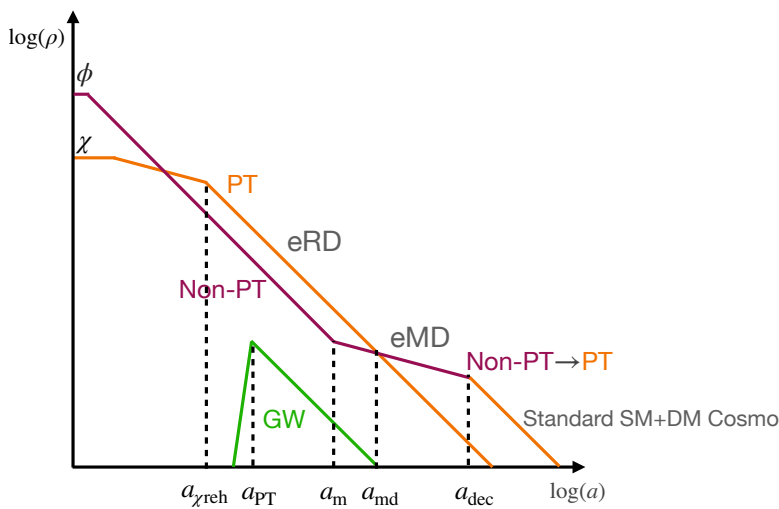}
		\caption{A schematic of the modified eMD model discussed in \Sec{sec:emdmodel} showing energy densities in various sectors as a function of the scale factor (time).
        While the decay products of the ALP $\chi$ are always radiation dominated, the massive particles (reheaton) in the decay products of the inflaton lead to a period of early matter domination.} 
		\label{fig:modified-ghost}
\end{figure}
Smaller $\f{PT}$ at $t_{\rm dec}$ allows for larger $\zeta_{\chi}$ while evading constraints from CMB summarized in \Eq{eq:fptconstraint}. If both PT and non-PT sectors are radiation-dominated throughout, then this also constrains the PT sector to be subdominant by the same factor $\f{PT}$ at the time of the phase transition, heavily suppressing the strength of GW produced, as seen in \Eqs{eq:ptgwestimateGhost}{eq:delomegagwPTghost}. 
This, however, will not be the case if there is a deviation from standard radiation dominance, for example, a period of early matter domination (eMD). 
In such modified cosmological histories, $\f{PT}$ is effectively a function of time, and can be such that the PT sector is dominant during the phase transition and only becomes subdominant by $t_{\rm dec}$.
We will show that this significantly enhances the strength of GWB compared to the purely-radiation-dominance scenario of \cite{Geller:2018mwu}, while still being consistent with constraints from CMB.
For clarity, we define this time dependent fraction as $\hat{f}(t)$, and its late time value at $t_{\rm dec}$ will continue to be denoted by $\f{PT}$.


Here we describe such a modified cosmological history with eMD, shown schematically in \Fig{fig:modified-ghost}. 
As in the minimal model, the inflaton reheats the non-PT sector at the end of inflation, which dilutes like radiation, $\rhox{nPT} \propto a^{-4}$.
On the other hand, $\rho_{\chi}$ acts like matter while oscillating around its minimum, which dilutes slower than radiation. 
If the oscillations persist for a sufficiently long  time, $\rho_{\chi}$ can dominate the energy density (as well as the curvature perturbations) of the universe before decaying to the PT sector, which subsequently undergoes the phase transition.
Notice that unlike the radiation-dominated model described previously, the PT sector dominates the energy density during the phase transition, and therefore $\rhox{GW}$ is not suppressed at production by $f_{\rm PT}$.
We will call this period of $\rhox{PT}$ domination the era of early radiation domination (eRD).

Now we take the non-PT sector to be dominated by massive particles which become non-relativistic around $t_{\rm m}$, sometime after the phase transition, redshifting slower than the radiation in the PT sector. The eRD of the PT sector is thereby replaced by eMD in the non-PT sector at $t_{\rm md}$. The period of eMD ends at $t_{\rm dec}$ when the non-PT massive particles decay, reheating the SM$+$DM. We will therefore refer to these  massive non-PT particles as ``reheatons''. 
The reheaton decays are taken to occur before DM decoupling, which allows thermalization with the SM and the remnants of the phase transition in the PT sector, matching onto the standard cosmological history from then on. In this way, we arrive at the following time-dependence of $\hat{f}_{\rm PT}$ from the $1/a^3$ redshifting of non-relativistic matter and the $1/a^4$ redshifting of relativistic radiation:
\begin{align}\label{eq:fpthat}
    \hat{f}_{\rm PT}(t) = \begin{cases} 
      \paren{1+\epsilon}^{-1} & \quad t_{\chi{\rm reh}}\leq t \leq t_m \\
      \paren{1+\epsilon \frac{a}{a_m}}^{-1} = \paren{1+\frac{a}{a_{\rm md}}}^{-1} & \quad t_m < t \leq t_{\rm dec} \,, \\
   \end{cases}
\end{align}
where $\epsilon \equiv \left.\paren{\frac{\rho_{\rm nPT}}{\rho_{\rm PT}}}\right|_{t_{\chi{\rm reh}}}$ and we have used $\epsilon (a_{\rm md}/a_m)=1$.
Therefore, $\f{PT} \equiv \hat{f}_{\rm PT}(t_{\rm dec}) = (1+a_{\rm dec}/a_{\rm md})^{-1}$.

\subsection{Strength of GW and anisotropies}
In this model with eMD, the PT sector dominates the energy density at the phase transition, $\rhox{total}(t_{\rm PT}) \approx \rhox{PT}(t_{\rm PT})$ and $\hat{f}_{\rm PT}(t_{\rm PT}) \approx 1$. 
Hence, in \Eq{eq:gwbubbles} for GW production at the time of the transition, $\f{PT}^{2}$  is not a suppression.

However, the period of eMD dilutes GW radiation because of the differential redshifting of matter and GW radiation, such that
\begin{align}
    \frac{\rhox{GW}(t_{\rm dec})}{\rhox{total}(t_{\rm dec})}   = \hat{f}_{\rm PT}(t_{\rm dec}) \frac{\rhox{GW}(t_{\rm PT})}{\rhox{total}(t_{\rm PT})}  \,.
\end{align}
This, along with the same microphysical parameters as in \Sec{sec:ghostgwstrength} for the phase transition, gives the fractional energy density in GWB today as 
\begin{align}\label{eq:ptgwestimateeMD}
    \Omgw 
    \approx & \, 3.2 \times 10^{-9} \f{PT}\,.
\end{align}
Note that the GW signal in this model, despite the dilution from eMD, has only one power of $\f{PT}$ suppression, compared to the quadratic suppression in \Eq{eq:gwPTghost}.


The anisotropy of the GWB can again be related to the fluctuations of $\chi$,
\begin{align}
    \delta_{\rm GW} \approx -\frac{4}{3} \zeta_{\chi} \,.
\end{align}
The factor in front of $\zeta_{\chi}$ is smaller compared to \Eq{eq:delgwghostsimplified} because the SW contribution, which also comes from the PT sector in this case, is anti-correlated with the inherent inhomogeneities of the GWB.
The expression for the anisotropies of the CMB for all relevant scales (i.e., the scales that re-enter much after $t_{\rm dec}$) remains the same as \Eq{eq:delcmbghost}.
This is shown explicitly in \Sec{sec:supermodes}. Then, even in the eMD model, we have
\begin{align}
    \delta_{\gamma} \approx  -\frac{4}{5} \zeta_{\phi} -\frac{4}{5} \f{PT} \zeta_{\chi} \,.
\end{align}
The inhomogeneity in the GWB is
\begin{align}
    \dOmgw \approx \Omgw \times \frac{4}{3}\zeta_{\chi}  \approx  4\times 10^{-9}   \f{PT} \zeta_{\chi} \,.
\end{align}
We must still satisfy the constraint in \Eq{eq:fptconstraint}. Saturating it gives
\begin{align}\label{eq:delomegagwPTeMD}
    \dOmgw  \leq 1.8 \times 10^{-13}  \,,
\end{align}
which is now independent of $\f{PT}$, as opposed to \Eq{eq:delomegagwPTghost}. 
The possibility that the relative anisotropies in the GWB are large while the absolute size of the anisoptropic signal is unsuppressed by $f_{\rm PT}$, is the  main result of our paper.
\subsection{Peak frequency of GWB}\label{sec:frequency}
The frequency spectrum of the GWB will be slightly shifted compared to the radiation-dominated scenario due to the modified cosmological history.
At production, the typical frequency of GWB for parameters given in \Sec{sec:ghostgwstrength} is \cite{caprini2016science}
\begin{align}
    \omega_{*} \approx 0.23 \tilde{\beta} H(t_{\rm PT}) .
\end{align}
Red-shifting this frequency gives the peak frequency of the spectrum today
\begin{align}
    \omega_{\rm peak} &=   \omega_{*} \paren{\frac{a_{\rm PT}}{a_0}}   \nonumber\\
    &\approx  0.23 \tilde{\beta} \paren{\frac{a_{\rm PT}}{a_0}} \paren{\frac{H(t_{\rm PT})}{H(t_{\rm md})}} \paren{\frac{H(t_{\rm md})}{H(t_{\rm dec})}} \paren{\frac{H(t_{\rm dec})}{H(t_{eq})}} \paren{\frac{H(t_{eq})}{H(t_0)}} H(t_0) .
\end{align}
Using $H \propto a^{-3/2}$ during matter dominance, while $H \propto a^{-2}$ during radiation dominance,
\begin{align}
    \omega_{\rm peak} \approx & \, 0.23 \tilde{\beta} \paren{\frac{a_{\rm PT}}{a_0}} \paren{\frac{a_{\rm md}}{a_{PT}}}^{2} \paren{\frac{a_{\rm dec}}{a_{\rm md}}}^{3/2} \paren{\frac{a_{eq}}{a_{\rm dec}}}^{2} \paren{\frac{a_0}{a_{eq}}}^{3/2} H(t_{0})\nonumber\\
    \approx & \, 2.8 \times 10^{-17} \tilde{\beta} \paren{\frac{a_{eq}}{a_{\rm PT}}} \f{PT}^{1/2} \,\,{\rm Hz} ,
\end{align}
where we have used the fact that the redshift at matter-radiation equality in the standard cosmological history $z_{eq} \sim a_{eq}^{-1} \sim 3100$,
$H(t_0) \sim 2.2 \times 10^{-18}$ Hz, and $a_{\rm md}/a_{\rm dec} \approx \f{PT}$ from \Eq{eq:fpthat}.
Now,
\begin{align}
    \frac{a_{eq}}{a_{\rm PT}} = \frac{T_{\rm PT}(t_{\rm PT})}{T_{\rm SM}(t_{eq}))} \f{PT}^{-1/4} 
    \paren{\frac{g_{*}(t_{\rm md})}{g_{*}(t_{\rm dec})}}^{1/4}.
\end{align}
Then with $T_{\rm SM}(t_{eq}) \sim 1$ eV,
$g_{*}(t_{\rm md}) \sim g_{*}(t_{\rm dec}) \sim 100$, 
we get
\begin{align}
    \omega_{\rm peak} \approx 2.8\times 10^{-5} \tilde{\beta} \f{PT}^{1/4} \paren{\frac{T_{\rm PT}(t_{\rm PT})}{\rm TeV}} \,{\rm Hz}.
\end{align}
Note that the dependence on $\f{PT}$ is very mild, less than an order of magnitude in the range that we will be led to consider.
For $\tilde{\beta} = 10 $ and $\f{PT} = 4.5 \times 10^{-3}$, a phase transition at $T_{\rm PT}(t_{\rm PT}) \approx 69$ TeV gives $\omega_{\rm peak} \approx 5$ mHz, which falls within the sensitivity of LISA \cite{amaro2017laser}, BBO \cite{Harry:2006fi}, and DECIGO \cite{Kawamura:2011zz}.
Choosing $T_{\rm PT}(t_{\rm md}) \approx 5.8$ TeV and $T_{\rm SM}(t_{\rm dec}) \approx 100$ GeV gives the correct duration of eMD.
For larger $\f{PT} = 4.5\times10^{-2}$, a smaller $T_{\rm PT}(t_{\rm PT}) \approx 39$ TeV gives the same peak frequency.
In this case, $T_{\rm PT}(t_{\rm md}) \approx 1$ TeV and $T_{\rm SM}(t_{\rm dec}) \approx 100$ GeV produces required duration of eMD.
Our choice of $T_{\rm SM}(t_{\rm dec}) \approx 100$ GeV here is motivated by the considerations discussed in \Sec{sec:smallstructure}. In principle, this temperature could be smaller depending on the model of DM.

\section{Gravitational crosstalk}\label{sec:gwcrosstalk}
The PT and non-PT sectors can influence each other gravitationally even though they are otherwise only weakly coupled.
We want to ensure that the CMB scales are unaffected despite large perturbations in the PT sector during eRD era.
We also want to identify other constraints on our eMD model which may arise from the evolution of perturbations after horizon re-entry.
We address these two concerns here.
\subsection{Modes that are superhorizon prior to $t_{\rm dec}$ }\label{sec:supermodes}
\begin{figure}[ht]
   \centering
   \includegraphics[width=0.6\linewidth]{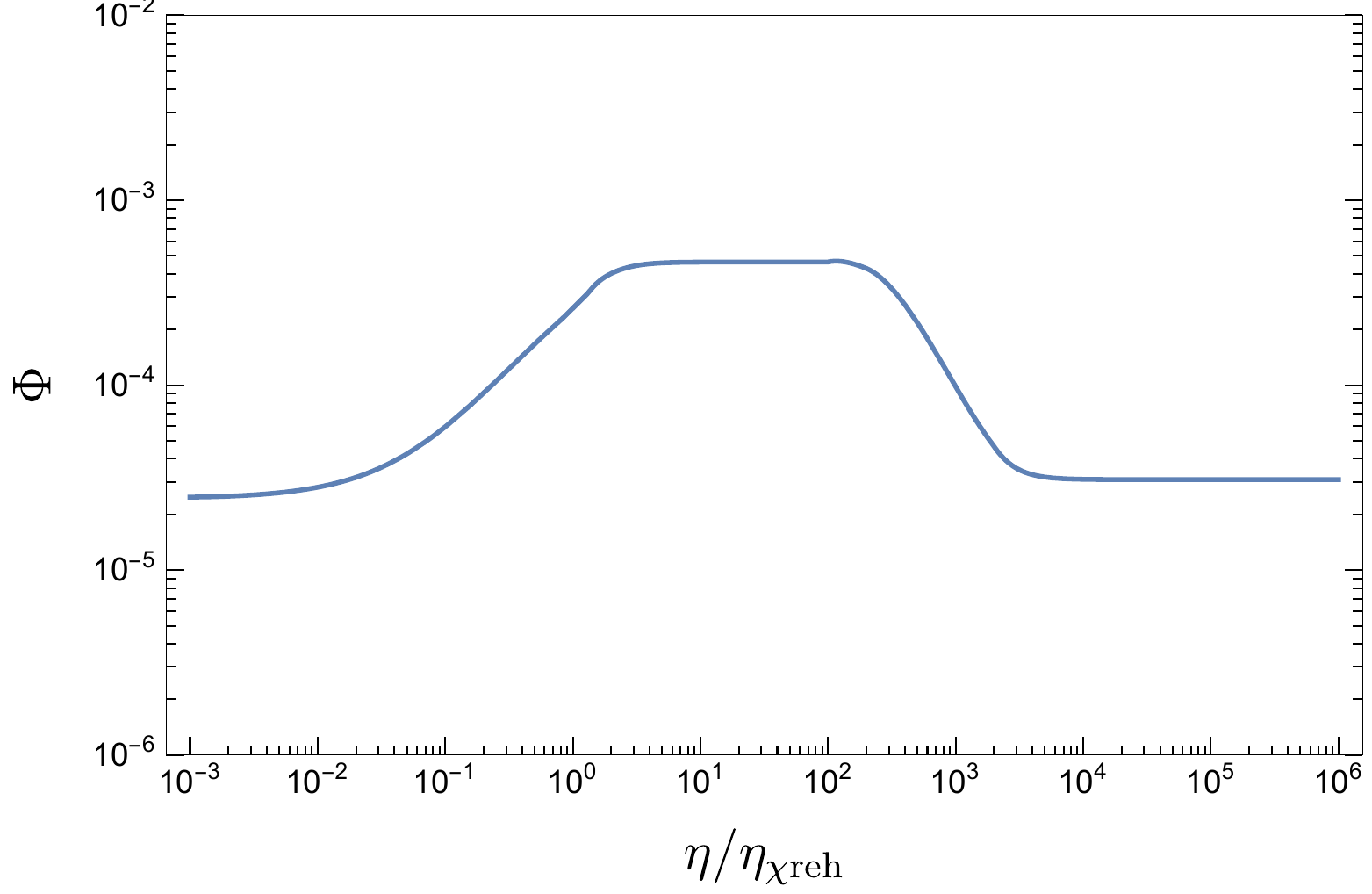}
	\caption{$\Phi(k)$ as a function of conformal time $\eta$ for a co-moving mode $k$ that re-enters sufficiently after the end of eMD. 
    We have taken a benchmark, $\zeta_{\chi} \approx 10^{-3}$, $\f{PT} \approx 10^{-2}$, and $\epsilon =1/2$.
    Conformal times denoting boundaries of different eras are 
    $\eta_{\rm m}  \sim \eta_{\rm md} \sim 100 \eta_{\chi {\rm reh}}$ and $\eta_{\rm dec} \sim 2000 \eta_{\chi {\rm reh}}$.} 
	\label{fig:Phisuper}
\end{figure}
Our modified early cosmology reverts to standard cosmology at $t_{\rm dec}$, where we will take $T_{\rm SM}(t_{\rm dec}) \gtrsim 100$ GeV.  We want to ensure that the observed fluctuation modes of the CMB and LSS would be unaffected by the modified cosmology before $t_{\rm dec}$, 
and this is relatively straightforward to analyze because they are all superhorizon prior to $t_{\rm dec}$, given the high $T_{\rm SM}(t_{\rm dec})$.

We use the convention where the perturbed FLRW metric is given by \cite{dodelson2020modern},
\begin{align}
    g_{00} = -1-2\Psi ,\quad g_{ij} = a^2 (t) \delta_{ij} [1+2\Phi] ,\quad g_{0i} =0 \,.
\end{align}
In the absence of anisotropic stress, $\Phi = -\Psi$, and the relevant Einstein equation is
\begin{align}\label{eq:Phifull}
    \eta \Phi'+ \Phi  + \frac{k^2 \eta^{2}}{3} \Phi & =  \frac{1}{2} \sum_{i} f_i \delta_{i}  \, ,
\end{align}
where $'$ denotes partial derivative with respect to the conformal time $\eta$ and $k$ is the co-moving momentum.  
$f_i \equiv \rho_{i}/\rho_{\rm total}$ is the fraction of the energy density in the $i^{\rm th}$ sector and $\delta_{i} \equiv \delta \rho_{i}/\rho_{i}$ is the fractional density perturbation.
When a certain mode is superhorizon, the $k^2$-term in \Eq{eq:Phifull} can be neglected, giving
\begin{align}\label{eq:Phisuper}
    \eta \Phi'+\Phi = \frac{1}{2} \sum_{i} f_i \delta_{i} \,.
\end{align}
The gauge invariant fluctuation of a non-interacting fluid is conserved on superhorizon scales. For fluid $i$, it is given by 
\begin{align}
    \zeta_{i} = \Phi +\frac{\delta_{i}}{3(1+w_i)} \,,
\end{align}
where $w_{i}$ is its equation of state.
Then $\zeta_{\rm nPT} = \zeta_{\phi}$ and $\zeta_{\rm PT} = \zeta_{\chi}$ remain constant until horizon re-entry. 
This also dictates the superhorizon evolution of radiation and matter density perturbations,
\begin{align}\label{eq:matsuper}
    \delta'_{\rm rad} = -4 \Phi' , \quad \delta'_{\rm mat} = -3 \Phi' \,.
\end{align}



    

Using \Eqs{eq:Phisuper}{eq:matsuper}, we numerically evaluated the evolution of $\Phi$ for $\zeta_{\chi} \approx  10^{-3}$ and $\f{PT} \approx 10^{-2}$, shown in \Fig{fig:Phisuper}.
While $\Phi$ is larger during eRD when the total curvature perturbation is dominated by $\zeta_{\chi}$, it becomes small again after eMD following $\zeta_{\phi}$\footnote{A similar conclusion was found in \cite{Suyama:2011pu}. They consider a model with two curvatons where one of the curvatons with larger perturbations dominates the energy density during an earlier era of radiation domination.}.
The fluctuation in the CMB can be written in terms of $\zeta_{\phi}$ and $\Phi$ as $\delta_{\gamma} \approx 4(\zeta_{\phi} + \f{PT} \zeta_{\chi} -2\Phi)$. 
We have chosen $\zeta_{\phi}$ such that $(\zeta_{\phi} + \f{PT} \zeta_{\chi}) \approx 4.5 \times 10^{-5}$.
From \Fig{fig:Phisuper}, $\Phi \approx 3\times 10^{-5}$ for the superhorizon modes during eRD. 
After matter-radiation equality, this will drop by a factor of ($9/10$) such that  $\Phi \approx 2.7 \times 10^{-5}$.
Then $\delta_{\gamma} \approx 4(\zeta_{\phi} + \f{PT} \zeta_{\chi} -2\Phi) \approx 3.6 \times 10^{-5}$, which is consistent with the observed value in the CMB.
This shows that the anisotropy in CMB (and in other maps of adiabatic perturbations) remains small at $10^{-5}$ even in our modified scenario with PT sector domination during eRD.

\subsection{Modes that have re-entered the horizon prior to $t_{\rm dec}$ }\label{sec:submodes}

In this section, we focus on the evolution of perturbations for modes that re-enter the horizon during eRD and eMD. 
This will be relevant for the constraints on $\zeta_{\chi}$ on small scales, which can indirectly affect CMB or LSS.

The gradient terms that we neglected for superhorizon evolution become important as the mode becomes subhorizon.
The evolution of $\Phi$ is 
\begin{equation}\label{eq:pertsub}
\begin{aligned}
    \eta \Phi'+ \Phi  + \frac{k^2 \eta^{2}}{3} \Phi & =  \frac{1}{2} \sum_{i} f_i \delta_{i}  \\
    \delta''_{\rm rad}+\frac{k^3}{3} \delta_{\rm rad} & = -4 \Phi'' + \frac{4 k^2}{3} \Phi  \\
     \delta''_{\rm mat} +\frac{1}{\eta} \delta'_{\rm mat} & = -3 \Phi'' +k^2 \Phi -\frac{3}{\eta} \Phi' \,.
\end{aligned}
\end{equation}
Consider a co-moving mode re-entering during eRD, $k_{\rm eRD}$.
Just before its horizon re-entry, 
$\Phi(k_{\rm eRD} \eta \lesssim 1) \approx (2/3) \zeta_{\chi}$ since the PT sector is dominant.
Now $\zeta_{\rm nPT} = \Phi + \delta_{\rm nPT}/4 \sim 10^{-5}$ is conserved on superhorizon scales.
Then for $\zeta_{\chi} \gg \zeta_{\phi}$, $\delta_{\rm nPT} (k_{\rm eRD} \eta <1) \approx -4\Phi \approx -(8/3) \zeta_{\chi}$ using \Eqs{eq:Phisuper}{eq:matsuper}.

After horizon re-entry at $\eta_{\rm eRD} \sim 1/ k_{\rm eRD}$, perturbations follow  eqs.~\ref{eq:pertsub}.
Below, we describe their qualitative behavior.
The fluctuations in the radiation of the PT sector, $\delta_{\rm PT}$, oscillates in time with a constant amplitude throughout eRD and eMD.
The behaviour of $\Phi$ and $\delta_{\rm nPT}$, however, changes from eRD to eMD.
During eRD, $|\Phi| \propto \eta^{-2}$ decreases, while the amplitude of $\delta_{\rm nPT}$ remains constant when it is radiation dominated and increases logarithmically $\sim \log(a)$ when it becomes non-relativistic. 
This logarithmic growth is insignificant if $\epsilon$ is close to 1, which we take to be the case.
The linear growth during eMD, however, is significant  when $\delta_{\rm nPT} \propto a$, and consequently $\Phi$ becomes constant.
In summary,  
\begin{equation}\label{eq:deltanPTeRD}
    \begin{aligned}
        |\delta_{\rm nPT}(k_{\rm eRD})| & \approx
        \begin{cases}
            \dfrac{8}{3} \zeta_{\chi} &  \eta_{\rm eRD} \leq \eta < \eta_{\rm md}  \vspace{0.3em} \\
            2 \dfrac{\eta^2}{\eta_{\rm md}^{2}} \zeta_{\chi} & \eta_{\rm md} \leq \eta < \eta_{\rm dec} \vspace{0.3em} \\
            \dfrac{8}{3} \dfrac{\eta_{\rm dec}^{2}}{\eta_{\rm md}^{2}} \zeta_{\chi} & \eta_{\rm dec} \leq \eta \,,
        \end{cases}
    \end{aligned}
\end{equation}
where we have used adiabatic relation $\delta_{\rm rad}/4 = \delta_{\rm mat}/3$ at the transitions between radiation to matter dominance in the non-PT sector, and $a \propto \eta^2$ during MD.
The evolution of $\Phi$ is
\begin{equation}\label{eq:Phierd}
    \begin{aligned}
        |\Phi(k_{\rm eRD})| &\approx 
    \begin{cases}
       \dfrac{2}{3} \dfrac{\eta_{\rm eRD}^{2}}{ \eta^{2}} \zeta_{\chi} &  \eta_{\rm eRD} \leq \eta < \eta_{\rm md} \vspace{0.3em}  \\
       \dfrac{3}{5} \dfrac{\eta_{\rm eRD}^{2}}{ \eta_{\rm md}^{2}} \zeta_{\chi} & \eta_{\rm md} \leq \eta < \eta_{\rm dec} \vspace{0.3em}  \\
       \dfrac{2}{3} \dfrac{\eta_{\rm eRD}^{2} \eta_{\rm dec}^{2}}{ \eta_{\rm md}^{2} \eta^2} \zeta_{\chi} & \eta_{\rm dec} \leq \eta \,,
    \end{cases} 
    \end{aligned}
\end{equation}
where we have used the fact that $\Phi$ decreases by a factor of ($9/10$) when the dominant density perturbations transition from RD to MD, and increases by ($10/9$) during the transition from MD to RD.

Now let us similarly analyse the evolution of perturbations for a co-moving mode $k_{\rm eMD}$ that re-enters during eMD. 
Since energy density in the PT sector with large perturbations drops during eMD, $\Phi(k_{\rm eMD})$ decreases even when the mode is superhorizon, as can be seen in \Fig{fig:Phisuper}. 
Using superhorizon conservation of $\zeta_{\rm nPT}$, we see that $\delta_{\rm nPT} \sim -\Phi$ also decreases.
After horizon re-entry, however, $\delta_{\rm nPT} \propto a$ increases while $\Phi$ remains constant till the end of the eMD era. Then
\begin{equation}\label{eq:deltanPTeMD}
    \begin{aligned}
        |\delta_{\rm nPT}(k_{\rm eMD})| & \approx
        \begin{cases}
            2 \dfrac{(\eta_{\rm md} \eta)^2}{\eta_{\rm eMD}^{4}} \zeta_{\chi} & \eta_{\rm eMD} \leq \eta < \eta_{\rm dec} \vspace{0.3em} \\
            \dfrac{8}{3} \dfrac{(\eta_{\rm md} \eta_{\rm dec})^2}{\eta_{\rm eMD}^{4}} \zeta_{\chi} & \eta_{\rm dec} \leq \eta \,,
        \end{cases}
    \end{aligned}
\end{equation}
and the evolution of $\Phi$ is
\begin{equation}\label{eq:Phiemd}
    \begin{aligned}
        |\Phi(k_{\rm eMD})| &\approx 
    \begin{cases}
       \dfrac{4 \hat{f}_{\rm PT}}{5 +\hat{f}_{\rm PT}} \zeta_{\chi} + \dfrac{3 (1-\hat{f}_{\rm PT})}{5 +\hat{f}_{\rm PT}} \zeta_{\phi} \approx \dfrac{4\eta_{\rm md}^{2}}{5\eta_{\rm eMD}^{2}} \zeta_{\chi} + \dfrac{3}{5} \zeta_{\phi} & \eta_{\rm eMD} \leq \eta < \eta_{\rm dec} \vspace{0.3em}  \\
        \dfrac{\eta_{\rm dec}^{2}}{\eta^{2}} \left[\dfrac{4 f_{\rm PT}}{5 } \zeta_{\chi} + \dfrac{4}{5} \zeta_{\phi}\right] \approx \dfrac{\eta_{\rm dec}^{2}}{\eta^{2}} \paren{\dfrac{4}{5} \dfrac{\eta_{\rm md}^{2}}{\eta_{\rm eMD}^{2}} \zeta_{\chi} + \dfrac{4}{5} \zeta_{\phi} } & \eta_{\rm dec} \leq \eta \,.
    \end{cases} 
    \end{aligned}
\end{equation}
Here, we have used $\f{PT} \ll 1$. The results of this section will be useful while analysing constraints on our eMD model, which we discuss next.



\subsubsection{Structure on small scales}\label{sec:smallstructure}
The density perturbations in the non-PT sector grow during eMD as seen in \Eqs{eq:deltanPTeRD}{eq:deltanPTeMD}.
These overdensities will be retained and transferred to the SM+DM plasma after the reheaton decay. 
This poses the danger of an overabundance of small scale structures like compact mini-halos in the late universe.
In reality, however, various processes at play during the decay of reheaton and DM decoupling suppress these small scale fluctuations, effectively erasing any accumulated growth from the eMD era. 
We list the relevant processes below:
\begin{enumerate}
    \item Refs.~\cite{Erickcek:2011us,Fan:2014zua} showed that the perturbative decay of the reheaton into radiation suppresses the amplitude of radiation perturbations by as much as $\sim 10^{-3}$ for subhorizon modes $k \gtrsim 20 k_{\rm dec}$.
    \item 
    Frictional damping and acoustic oscillations during kinetic decoupling, and the free-streaming of DM afterwards erases DM fluctuations on all length scales smaller than the maximum of the horizon size at decoupling and free-streaming length-scale \cite{Loeb:2005pm, Bertschinger:2006nq}.
    These effects erase enhancement in DM perturbations for all modes $k \gtrsim k_{\rm dec}$ that re-entered during eRD and eMD.
    \item Finally, large perturbations in the baryon and photon plasma are erased by Silk damping \cite{silk1968cosmic,hu1996small}, and thus do not contribute to additional small-scale structure. 
\end{enumerate}
These 
mechanisms that erase DM perturbations on small scales require DM to be in thermal equilibrium with the SM at the time of reheaton decay.
Our choice of $T_{\rm SM}(t_{\rm dec}) \gtrsim 100$ GeV ensures that a standard WIMP-like DM candidate would satisfy this criterion.
In summary, enhanced structure formation on small scales can be easily avoided in our eMD model by taking appropriately high reheat temperature for the SM at $t_{\rm dec}$.

\subsubsection{Primordial Black Holes}\label{sec:pbh}
Horizon patches with sufficient overdensity can collapse to form primordial black holes (PBH) in the early universe, which can not be erased by the mechansims discussed above.
PBH production from adiabatic perturbations $\sim 10^{-5}$ is negligible.
However, in our modified model, larger density perturbations during eRD and eMD era will enhance PBH production, giving a constraint on the power spectrum of $\chi$ on small scales.


During radiation dominance, a co-moving mode $k$ contributes to PBH formation dominantly at its horizon re-entry.
The mass of the corresponding PBH ($M(k)$) is roughly the horizon mass at that time \cite{Green:1997sz},
\begin{align}\label{eq:pbhmass}
    M(k) = \gamma M_{H}(k\sim aH) \sim  \left. 0.2 \times \frac{4\pi}{3} \rho_{\rm total} H^{-3} \right|_{k=aH}
\end{align}
where we have taken $\gamma \sim 0.2$.
Only the regions with density greater than the critical overdensity $\delta_{c}$ collapse to form PBH.
For Gaussian perturbations, the probability of collapse during RD is given by \cite{green2021primordial}
\begin{align}\label{eq:betarad}
    \beta_{\rm rad}(M) \approx \sqrt{\frac{2}{\pi}} \gamma \frac{\sigma(M)}{\delta_c} \, \exp\paren{-\dfrac{\delta_{c}^{2}}{2 \sigma^{2}(M)}} \,.
\end{align}
Here $\sigma^{2}(M)$ is the mass variance, which is related to the power spectrum of the curvature perturbations ${\cal P}_{\cal R}$\footnote{The power spectrum for any gauge-invariant perturbation `$i$' is defined as
\begin{align}
     \< \zeta_{i}(\vec{k}) \zeta_{i}(\vec{k}') \> \equiv  \delta(\vec{k}+\vec{k}') \frac{(2\pi)^3}{k^3} {\cal P}_{i}(k) \, ,
\end{align}
For a scale-invariant spectrum, ${\cal P} (k) $ is constant.},
\begin{align}\label{eq:massvar}
    \sigma^{2}(M) \equiv \langle (\delta M / M)^2\rangle = \frac{16}{81} \int_{0}^{\infty} (q/k)^2 j_{1}^{2}(q/(\sqrt{3} k)) \tilde{W}^{2}(q/k) {\cal P}_{\cal R} \frac{dq}{q} \,.
\end{align}
Here, $j_{1}$ is the spherical Bessel function and  $\tilde{W}(q/k)$ is the Fourier transform of 
a window function used to smooth density contrast over the length scale $k^{-1} \sim (aH)^{-1}$.
During eRD, ${\cal P}_{\cal R} \approx {\cal P}_{\chi}$. A real-space top hat window function in \Eq{eq:massvar} gives $\sigma^2(M)  \sim 1.1 {\cal P}_{\chi}$ \cite{PhysRevD.97.103528}. 
We take $\delta_{c} \approx 0.4$ \cite{PhysRevD.100.123524}.
Evaporation of light PBH from eRD era can cause CMB distortions.
This gives a constraint on their formation probability $\beta_{\rm eRD} \lesssim 10^{-25}$ \cite{carr2021constraints,carr2022primordial}. 
Despite the strong constraint on $\beta_{\rm eRD}$, its exponential dependence on ${\cal P}_{\chi}$ from \Eqs{eq:betarad}{eq:massvar} translates to a rather weak constraint on $\zeta_{\chi} < 0.04$.
A more significant enhancement of PBH formation occurs during the eMD era, which we discuss below.

Density perturbations in the non-PT sector would grow linearly during MD, $\delta_{\rm nPT} \propto a(t)$, as discussed in \Sec{sec:submodes}.
Thus, even smaller perturbations can lead to significant PBH formation compared to the radiation era.
When the perturbation becomes non-linear $\delta_{\rm nPT} \gtrsim {\cal O}(1)$, the corresponding Hubble patch separates from the background expansion and collapses eventually.
It was, however, shown in refs.~\cite{khlopov1980primordial,Harada:2016mhb} that only patches with sufficient spherical symmetry and homogeneity can collapse to PBH\footnote{The patches that do not satisfy these criteria form pancake/cigar shaped structures instead. These are later dispersed due to the radiation pressure when the non-PT sector decays to SM.}.
The probability of PBH formation during matter domination is then
\begin{align}\label{eq:betamat}
    \beta_{\rm mat}(k) \approx 0.2 \paren{\dfrac{r_{\rm sch}}{r_{k}}}^{13/2}
\end{align}
where $r_{k} \sim a(t) k^{-1}$ is the physical wavelength of the mode and
$r_{\rm sch}$ is the Schwarzschild radius corresponding to the typical overdense region on length scale $r_{k}$.
The ratio $(r_{\rm sch}/r_{k}) \approx \delta_{\rm nPT}(k,t) (aH/k)^2$ remains constant during eMD due to the linear growth of $\delta_{\rm nPT}$. 
Hence, the ratio can be evaluated at horizon re-entry for the modes re-entering during eMD, while for modes that have re-entered before eMD, the ratio is evaluated at the start of eMD.
From \Eqs{eq:deltanPTeRD}{eq:deltanPTeMD}, we get
\begin{align}\label{eq:rratio}
    \frac{r_{\rm sch}}{r_{k}} \approx 
    \begin{cases}
        \delta_{\rm nPT} (k,a_{\rm md}) \paren{\dfrac{a_{k}}{a_{\rm md}}}^{2} \approx 2 \zeta_{\chi} \paren{\dfrac{a_{k}}{a_{\rm md}}}^{2} & k \geq k_{\rm md} \vspace{0.3em}\\
        \delta_{\rm nPT} (k,a_k) \approx 2 \zeta_{\chi} \paren{\dfrac{a_{\rm md}}{a_k}} &  k_{\rm md} > k > k_{\rm dec},        
    \end{cases}
\end{align}
where $a_k$ corresponds to the scale factor at the horizon re-entry of the $k$-mode, and we have used $H \propto a^{-2}$ during RD while  $H \propto a^{-3/2}$ during MD.
Clearly, the largest $\beta$ corresponds to the mode re-entering at the start of eMD, $k_{\rm md}$, with $\beta_{\rm mat}({k_{\rm md}}) \approx 18 \zeta_{\chi}^{13/2}$.
The mass relation given in \Eq{eq:pbhmass} also holds for PBH formed during eMD with only a difference of ${\cal O}(1)$ factor.
 

Modes re-entering during lRD have curvature perturbation $\sim 10^{-5}$ as shown in \Sec{sec:supermodes}, and produce negligible amount of PBH following \Eq{eq:betarad}.
The modes that re-entered earlier during eRD and eMD, however, retain larger perturbations even in the lRD era (till they are erased by machanisms discussed in \Sec{sec:smallstructure}). 
We will show that despite large perturbations, there is no additional enhancement in PBH formation from these modes in the lRD era. 
The collapse probablity of a subhorizon mode in lRD can be estimated by performing Jeans mass analysis.
Balancing kinetic and gravitational potential energy gives the Jeans mass $M_J = 2 r/G$ for a spherical region of radius $r$. (Here, $G$ is the gravitational constant.)
The condition for instability $M >M_J$ then gives the critical overdensity for subhorizon modes,
\begin{align}
    \delta_{c} = \frac{1}{2 G \rho r^2} = \frac{1}{2 \pi^2 G \rho} \frac{k^2}{a^2} \,,
\end{align}
where we have taken $r = \pi a/k$. 
Using $8\pi G \rho = 3 H^2$, the critical overdensity at some time  $t_{\rm eRD}$ during eRD is
\begin{align}
    \delta_{c} =  \frac{4}{3\pi}\paren{\frac{k^2}{(a_{\rm eRD} H_{\rm eRD})^2}} = \frac{4}{3\pi}\paren{\frac{(a_k H_k)^2}{(a_{\rm eRD} H_{\rm eRD})^2}} = 0.43 \paren{\frac{a_{\rm dec}}{a_k}} \paren{\frac{a_{\rm eRD}}{a_{\rm dec}}}^{2} \,,
\end{align}
Note that the critical overdensity is bigger for subhorizon modes than the constant $\approx 0.4$ we had taken before.
Remember that $\delta_{\rm nPT}(k,t_{\rm eRD}) = \delta_{\rm nPT}(k,t_{k})(a_{\rm dec}/a_k)$ due to the growth during eMD. 
Then the ratio $(\delta_{c}/\delta_{\rm nPT})$ is the smallest when $t_{\rm eRD} = t_{\rm dec}$ where
\begin{align}
    \dfrac{\delta_{c}(t_{\rm dec})}{\delta_{\rm nPT}(k,t_{\rm dec})} \approx \dfrac{0.43}{\delta_{\rm nPT}(k,t_{k})} \geq \dfrac{0.43}{\delta_{\rm nPT}(k_{\rm md}, t_{\rm md})} .
\end{align}
Using \Eqs{eq:betarad}{eq:massvar}, the probability of collapse of subhorizon modes during lRD is then
\begin{align}\label{eq:betalRD}
    \beta_{\rm lRD} \lesssim  \exp \paren{-\frac{0.43^2}{2(1.1 {\cal P}_{\chi})}} \,.
\end{align}
Following the analysis for eRD at the beginning of this section, we see that this production probability is also much smaller than that for the eMD era.



Having analysed PBH production, let us connect it to the experimental bounds. Constraints on PBH are typically cast in terms of its abundance compared to the DM today, $\f{PBH} \equiv \frac{\Omega^{(0)}_{\rm PBH}}{\Omega^{(0)}_{\rm DM}}$, with $\Omega^{(0)}_{\rm DM} \approx 0.26$.
At formation, $\rho_{\rm PBH}(M) \approx \beta(M) \rho_{\rm total}$, which then redshifts like matter till today.
$\rho_{\rm PBH}$ dilutes slower than $\rho_{\rm total}$ when the universe is radiation dominated.
This gives relative enhancement of $\Omega_{\rm PBH}$ during periods of radiation dominance.
The abundance of PBH today can be evaluated as
\begin{align}
    \Omega^{(0)}_{\rm PBH}(M) \sim 
    \begin{cases}
        \beta_{\rm rad}(M) \paren{\dfrac{a_{eq}}{a_M}} \f{PT}+\beta_{\rm mat}(M) \paren{\dfrac{a_{eq}}{a_{\rm dec}}} & a_M \leq a_{\rm md} \vspace{0.3em}\\
        \beta_{\rm mat}(M) \paren{\dfrac{a_{eq}}{a_{\rm dec}}} &  a_{\rm md} < a_M < a_{\rm dec} ,
    \end{cases}
\end{align}
where $a_{M}$ corresponds to the time of re-entry of the mode that produces PBH of mass $M$.

The PBH mass can be similarly related to $a_M$ using \Eq{eq:pbhmass} along with the relation $\rho_{\rm total} \approx 3 H^2 \mpl^{2}$, 
\begin{align}\label{eq:pbhMaM}
    M(a_{M}) \sim  
    \begin{cases}
    10^{-6} M_{\odot} \paren{\dfrac{a_M}{a_{\rm md}}}^{2} \f{PT}^{3/2} \paren{\dfrac{100 {\rm GeV}}{T_{\rm SM}(t_{\rm dec})}}^{2} & a_M \leq a_{\rm md} \vspace{0.3em}\\
    10^{-6} M_{\odot}  \paren{\dfrac{a_M}{a_{\rm dec}}}^{3/2}  \paren{\dfrac{100 {\rm GeV}}{T_{\rm SM}(t_{\rm dec})}}^{2} &  a_{\rm md} < a_M < a_{\rm dec}
    \end{cases}    
\end{align}
Here, $M_{\odot}$ corresponds to the solar mass, and we have used $\mpl \sim 10^{-38} M_{\odot}$.

\begin{figure}[t]
   \centering
   \includegraphics[width=0.7\linewidth,trim={2.5cm 4.5cm 6cm 5.cm},clip]{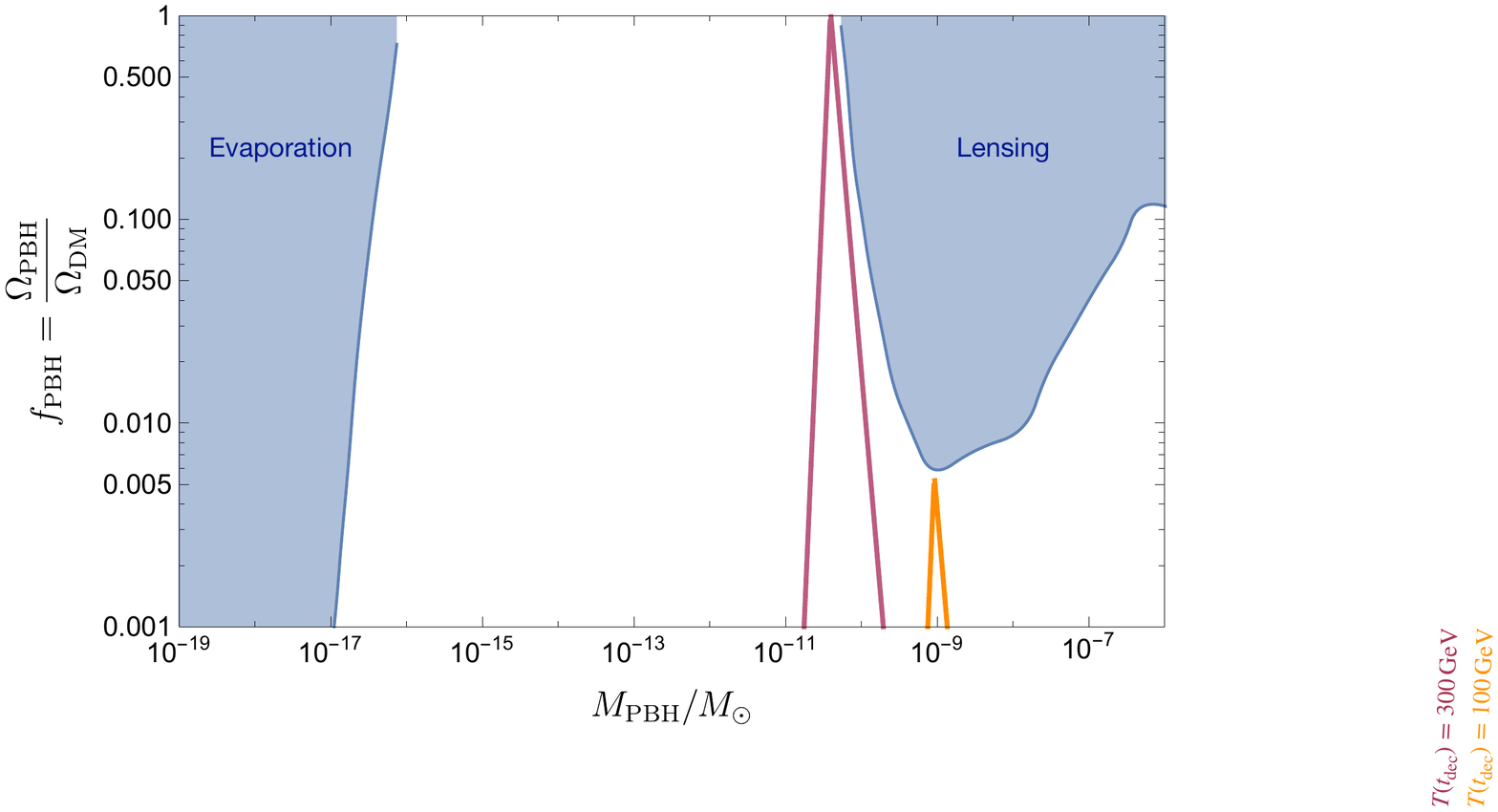}
	\caption{Fraction of DM energy density in PBH as a function of PBH mass in units of solar mass $M_{\odot}$. The blue regions show current constraints from CMB distortions caused by light evaporating PBH and gravitation lensing constraints at higher $M$ (taken from \cite{carr2021constraints,carr2022primordial}). We show PBH abundances for two benchmarks in the eMD model: (orange) $\zeta_{\chi} \approx 5 \times 10^{-3}$, $\f{PT} \approx 9 \times 10^{-3}$, $T_{\rm SM}(t_{\rm dec}) = 100 $ GeV, and (purple) $\zeta_{\chi} \approx 10^{-2}$, $\f{PT} \approx 4.5 \times 10^{-3}$, $T_{\rm SM}(t_{\rm dec}) = 300$ GeV. We see that in the second case, PBH could make up the entirety of DM.}
	\label{fig:pbhvsM}
\end{figure}
Fig.~\ref{fig:pbhvsM} shows $\f{PBH}$ as a function of PBH mass compared to the experimental constraints \cite{carr2021constraints,carr2022primordial} shown as blue regions. 
We have chosen two benchmarks maximising the constraint in \Eq{eq:fptconstraint} and assuming a scale-invariant spectrum: 
\begin{enumerate}
    \item  $\zeta_{\chi} \approx 5\times 10^{-3}$, $\f{PT} \approx 9 \times 10^{-3}$, and $T_{\rm SM}(t_{\rm dec}) =100$ GeV (orange).
    \item  $\zeta_{\chi} \approx 10^{-2}$, $\f{PT} \approx 4.5 \times 10^{-3}$, and $T_{\rm SM}(t_{\rm dec}) =300$ GeV (purple).
\end{enumerate}
Our eMD model effectively produces mono-chromatic PBH with the peak corresponding to PBH produced from the co-moving mode  $k_{\rm md}$ that re-enters the horizon at $t_{\rm md}$.
We see that for the second benchmark, PBH can make up the entirety of DM.
This however requires higher SM temperature at $t_{\rm dec}$, which in turn requires higher temperature in the PT sector at the start of eMD, $T_{\rm PT}(t_{\rm md}) = 18$ TeV. 
The analysis of \Sec{sec:frequency} then shows that in this case, the eMD must start soon after the phase transition. 
A larger PBH fraction can also be obtained at lower $T_{\rm SM}(t_{\rm dec})$ by taking smaller $\f{PT}$ (and not saturating \Eq{eq:fptconstraint}).
Generally, $\zeta_{\chi} \lesssim 10^{-2}$ is required to avoid overproduction of PBH.
The exact bound depends on the values of $T_{\rm SM}(t_{\rm dec})$ and $\f{PT}$ as they can shift the location of the peak, exposing it to stricter constraints. 
However, \Eqs{eq:betamat}{eq:rratio} suggest that even a moderately smaller $\zeta_{\chi}$ will be sufficient to evade these constraints.

In conclusion, we require $\zeta_{\chi} \lesssim 10^{-2}$ on small scales to avoid overproduction of PBH.
This also constrains $\zeta_{\chi}$ and $\delta_{\rm GW}$ on large scales if the power spectrum of $\chi$ is nearly scale-invariant.
A bigger $\zeta_{\chi}$ on large scale can be made consistent with PBH constraints if the power spectrum is significantly red-tilted.
However, this does not happen readily for a light field, and would require special model-building for $\chi$ potential, which we do not consider in this paper.  


 

\subsubsection{Scalar-induced gravitational waves}
A weaker gravitational wave background can be induced by curvature perturbations at second order in perturbation theory (see \cite{Domenech:2021ztg} for a recent review). 
Larger curvature perturbations and the presence of eMD can enhance its production.
Here, we estimate the strength of such induced-GWB and compare it to the gravitational waves from the phase transition.

Fraction of energy density in the induced-GWB per logarithmic $k$ is
\begin{align}\label{eq:indGWexpression}
    \Omega_{\rm GW}^{\rm ind} (\eta, k)= \frac{1}{24} \paren{\frac{k}{a(\eta)H(\eta)}}^2 \overline{{\cal P}_{h}(\eta, k)} 
\end{align}
where $\overline{{\cal P}_{h}(\eta, k)}$ is the time-averaged tensor power spectrum.
It is related to the power spectrum of curvature perturbations as
\begin{align}
    {\cal P}_{h}(\eta, k) = 4  \int_{0}^{\infty} dv \int_{|1-v|}^{|1+v|} du \left[ \frac{4 v^2 - (1+v^2-u^2)^2}{4vu} \right]^2 [I(v,u,x)]^2 {\cal P}_{\cal R}(kv) {\cal P}_{\cal R}(ku) 
\end{align}
where  $u = |\vec{k}-\vec{k'}|/k$, $v = k'/k$, and $x\equiv k \eta$. $I(v,u,x)$ contains information about the dynamics of scalar and tensor perturbations and is given in Ref.~\cite{Kohri:2018awv}. 
The curvature perturbations can be related to (first-order) $\Phi$ as 
\begin{align}
    {\cal P}_{\cal R} = \paren{\frac{5+3w}{3+3 w}}^{2} {\cal P}_{\Phi}
\end{align}
Thus, the magnitude of ${\cal P}_{h} \propto {\cal P}_{\Phi}^{2}$.

During radiation dominance, $\Phi(\eta,k) \propto 1/\eta^2$ decays quickly inside the horizon. 
Thus, induced-GW are dominantly produced at horizon re-entry for modes that re-enter deep in the eRD era.
This is not the case for modes re-entering during eMD, 
since $\Phi$ remains constant even after horizon re-entry during eMD.
The result is a relative enhancement of induced-GW during eMD compared to the case of pure radiation dominance.

\begin{figure}[t]
   \centering
   \includegraphics[width=0.7\linewidth,trim={2.5cm 5.3cm 6cm 5.cm},clip]{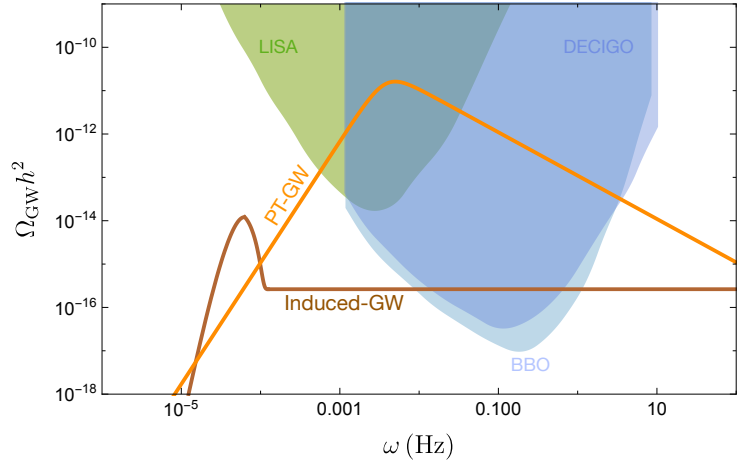}
	\caption{Strength of GWB from phase transition (orange) and induced from curvature perturbations at second order (brown), compared with the power-law integrated sensitivities of LISA, BBO, and DECIGO \cite{Schmitz:2020syl}.
    The GWBs are computed for the benchmark that saturates PBH constraint (see discussion in \Sec{sec:pbh}) with $\zeta_{\chi} = 10^{-2}$, $\f{PT} = 4.5 \times 10^{-3}$, and $T_{\rm SM}(t_{\rm dec}) = 300$ TeV.} 
	\label{fig:gwvsf}
\end{figure}
Refs.~\cite{Baumann:2007zm,Assadullahi:2009nf,Kohri:2018awv} have analysed production of induced-GW in matter dominance era for a scale-invariant ${\cal P}_{\Phi}$ that remains constant on super-horizon scales\footnote{It was pointed out in \cite{Inomata:2019ivs} and \cite{Inomata:2019zqy} that the strength of the induced GW from eMD sensitively depends on the exact model for the decay of the matter at the end of eMD. 
It was observed that a slow decay process suppresses the induced GW signal compared to \Eq{eq:indgw}, while a quick decay enhances the signal. 
We suspect that a typical perturbative decay of the reheaton will fall under `slow decay', suppressing induced GW signal. However, we have taken a conservative estimate following \cite{Kohri:2018awv}. 
}.
We borrow results of \cite{Kohri:2018awv} with appropriate rescaling as explained below.
While $\Phi(k) \sim \zeta_{\chi}$ remains constant for a superhorizon mode during eRD,
it decreases during eMD till horizon re-entry as 
the PT sector becomes subdominant (see \Fig{fig:Phisuper}).
After horizon re-entry, $\Phi(k)$ becomes constant again till the end of eMD, and this value depends on the time of re-entry for each mode.
The dominant effect of this evolution of $\Phi$ on the strength of induced-GW is the $k$-dependent rescaling of the results in \cite{Kohri:2018awv} by a factor $\lambda(k)$.
This rescaling factor $\lambda(k)$ can be obtained by noting that ${\cal P}_{h} \propto {\cal P}_{\Phi}^{2}$, and using \Eq{eq:Phiemd},
\begin{align}
    \lambda(k) \approx 
    \left[\dfrac{3}{2}\paren{\dfrac{4 \hat{f}_{\rm PT}(t_k)}{5+\hat{f}_{\rm PT}(t_k)}}\right]^4 
\end{align}
Then the strength of the induced-GW is given as
\begin{align}\label{eq:indgw}
     \frac{\Omega_{\rm GW}^{{\rm sec} (0)} (k) }{{\cal P}_{\chi}^{2} \Omega_{\rm rad}^{(0)}} \approx 
     \begin{cases}
      0.8 \f{PT} & k > 2 k_{\rm md} \vspace{0.3em}\\
        0.8 \f{PT} + \left(\dfrac{\lambda(k) }{14000} \paren{\dfrac{k}{k_{\rm dec}}}^2 (1-2\tilde{k}^{-1})^4 \right.\vspace{0.2em}\\
        \hspace{4em} \left. \dfrac{}{}\times  \left[105 \tilde{k}^{2}+72 \tilde{k} +16 -32 \tilde{k}^{-1}-16 \tilde{k}^{-2}\right]\right)  &   2 k_{\rm md} \geq k \geq k_{\rm md} \vspace{0.3em}\\
         \dfrac{\lambda(k)}{14000} \paren{\dfrac{k}{k_{\rm dec}}}^2 \left[1792 \tilde{k}^{-1}-2520 + 768\tilde{k} -105 \tilde{k}^{2}\right]  &   k_{\rm md} > k > k_{\rm dec}
     \end{cases}
\end{align}
where $\tilde{k}\equiv k/k_{\rm md}$.
The frequency of the GW today can be related to the co-moving mode $k$ as
\begin{align}
    \omega = \frac{k}{2 \pi} \sim 1.5 \times 10^{-15} \paren{\frac{k}{1 {\rm Mpc}^{-1}}} {\rm Hz}
\end{align}

Figure~\ref{fig:gwvsf} shows $\Omega_{\rm GW}$ from phase transition (orange) and induced from curvature perturbations (brown) as a function of frequency. We have taken the largest $\zeta_{\chi} \approx 10^{-2}$ allowed on small scales from PBH consideration (see benchmark 2 from \Sec{sec:pbh}). 
We see that the GW signal from the phase transition is much larger than the induced-GW. Decreasing $\zeta_{\chi}$ will quickly make induced-GW signal undetectably small given its sensitive dependence on the curvature perturbations that can be seen in \Eq{eq:indgw}.
Interestingly, while a small $\f{PT}$ suppresses GW signal from the phase transition, it enhances induced-GW. We will explore this aspect of induced-GWB in future work.

\section{Benchmarks and Observability}\label{sec:benchmarks}
The energy density in any GWB is bounded by two considerations.
Firstly, it  contributes to $N_{\rm eff}$ as a species of dark radiation, but the $N_{\rm eff}$ constraint is relatively mild:
\begin{align}
    \rhox{GW} < 0.1 \rho_{\gamma}.
\end{align}
Since we are interested in the possibility of $\delta_{\rm GW} \gg 10^{-5}$, it necessitates an isocurvature origin for the GWB.  The stronger bounds on isocurvature from CMB give
\begin{align}
    \delta \rhox{GW} < 0.1 \delta \rho_{\gamma}.
\end{align}
This can be translated to 
\begin{align}
    \dOmgw  < 10^{-5} \Omega_{\gamma}^{(0)} h^2 \sim 2.5 \times 10^{-10},
\end{align}
where $\Omega_{\gamma}^{(0)} h^2 \approx 2.5 \times 10^{-5}$.
We choose benchmarks that satisfy this constraint and also show that 
our modified eMD scenario brings $\delta \Omega_{\rm GW}^{(0)}$ of the signal closer to this upper bound over a large range of anisotropies.

In table.~\ref{tab:benchmarks}, we consider benchmarks that saturate \Eq{eq:fptconstraint}, i.e. $\zeta_{\chi} \f{PT} = 4.5 \times 10^{-5}$.
Corresponding GWB signals and their inhomogeneities are computed for the radiation-dominated model of \cite{Geller:2018mwu} using (more precise) \Eq{eq:gwbubbles} for benchmark 1, and eqs.~\ref{eq:ptgwestimateGhost} and \ref{eq:delomegagwPTghost} for the rest.
The same quantities are also computed for our eMD scenario using eqs.~\ref{eq:ptgwestimateeMD} and \ref{eq:delomegagwPTeMD}.
We see that the isotropic GWB signal is larger in the eMD model for all $\zeta_{\chi} \gtrsim 10^{-4}$, while the absolute inhomogeneities are larger for $\zeta_{\chi} \gtrsim 10^{-3}$.
\begin{table}[h!]
    \centering
    \begin{tabular}{|c|c||c|c|c|c|} 
        \hline
       & & \multicolumn{2}{|c|}{{\rm Radiation-dominance}} & \multicolumn{2}{|c|}{{\rm eMD}} \\
        \hline 
        & & $\Omega_{\rm GW}^{(0)}$ & $\delta \Omega_{\rm GW}^{(0)}$ & $\Omega_{\rm GW}^{(0)}$ & $\delta \Omega_{\rm GW}^{(0)}$ \\ 
        \hline\hline
        1) &$\zeta_{\chi} \sim  10^{-4}$, $\f{PT} \sim  4.5\times 10^{-1}$  & $1.2\times 10^{-9}$ & $5\times 10^{-13} $ & $3.2\times 10^{-9}$ & $1.9\times 10^{-13} $ \\[1ex]
       2) &$\zeta_{\chi} \sim  10^{-3}$, $\f{PT} \sim 4.5\times 10^{-2}$ & $2.6\times 10^{-11}$ & $ 10^{-13} $ & $1.5\times 10^{-10}$ & $1.9\times 10^{-13} $\\[1ex]
       3) & $\zeta_{\chi} \sim 10^{-2}$, $\f{PT} \sim 4.5\times 10^{-3}$ & $2.6\times 10^{-13}$ & $ 10^{-14} $ & $1.5\times 10^{-11}$ & $1.9\times 10^{-13} $\\[1ex]
       \hline\hline
       4) &$\zeta_{\chi} \sim 10^{-1}$, $\f{PT} \sim 4.5 \times 10^{-4}$  & $2.6\times 10^{-15}$ & $ 10^{-15} $ & $1.5\times 10^{-12}$ & $1.9\times 10^{-13} $\\
       \hline
    \end{tabular}
    \caption{Benchmarks showing the strength of GWB and its inhomogeneities in the pure radiation-dominance model of \cite{Geller:2018mwu} and our eMD model. We take them to saturate the CMB bound in \Eq{eq:fptconstraint} (that is, $\zeta_{\chi} \f{PT} = 4.5 \times 10^{-5}$). For a scale-invariant spectrum, they are consistent with the constraints from PBH production discussed in \Sec{sec:pbh}, except for benchmark 4, which would require significantly red-tilted spectrum for $\chi$ to evade constraints on small scales (high-$k$).}
    \label{tab:benchmarks}
\end{table}

Note that the benchmark 4 apparently violates the overproduction of PBH bounds discussed in \Sec{sec:pbh}. However, that analysis assumed that 
$\zeta_{\chi}$ was approximately scale-invariant up to the high-$k$ modes relevant for PBH production. This is indeed the case for the minimal model we have studied of a light $\chi$ with fixed mass during inflation. But it is possible  that with non-minimal couplings,
$\zeta_{\chi}$ has a significant red tilt so that the PBH bounds are evaded while still having large $\delta_{\rm GW}$ on observable scales (small $k$).

\begin{table}[h]
    \centering
    \begin{tabular}{|c|c||c|c|c|c|} 
        \hline
       & & \multicolumn{2}{|c|}{{\rm Radiation-dominance}} & \multicolumn{2}{|c|}{{\rm eMD}} \\
        \hline 
        & & $\Omega_{\rm GW}^{(0)}$ & $\delta \Omega_{\rm GW}^{(0)}$ & $\Omega_{\rm GW}^{(0)}$ & $\delta \Omega_{\rm GW}^{(0)}$ \\ 
        \hline\hline
        1) &$\zeta_{\chi} \sim  10^{-4}$, $\f{PT} \sim 4.5 \times 10^{-2}$ & $2.6\times 10^{-11}$ & $ 10^{-14} $ & $1.5\times 10^{-10}$ & $1.9\times 10^{-14} $\\[1ex]
       2) &$\zeta_{\chi} \sim  10^{-3}$, $\f{PT} \sim 4.5 \times 10^{-3}$ & $2.6\times 10^{-13}$ & $ 10^{-15} $ & $1.5\times 10^{-11}$ & $1.9\times 10^{-14} $\\[1ex]
       3) & $\zeta_{\chi} \sim 10^{-2}$, $\f{PT} \sim 4.5 \times 10^{-4}$ & $2.6\times 10^{-15}$ & $ 10^{-16} $ & $1.5\times 10^{-12}$ & $1.9\times 10^{-14} $\\[1ex]
       \hline\hline
       4) &$\zeta_{\chi} \sim 10^{-1}$, $\f{PT} \sim 4.5 \times 10^{-5}$  & $2.6\times 10^{-17}$ & $ 10^{-17} $ & $1.5\times 10^{-13}$ & $1.9\times 10^{-14} $\\
       \hline
    \end{tabular}
    \caption{Benchmarks with the PT sector making up only about $10\%$ of the final SM+DM density perturbations, $\zeta_{\chi} \f{PT} = 4.5 \times 10^{-6}$. The strength of GWB and its inhomogeneities is shown for both radiation-dominance and eMD models. The difference between the two is greater compared to the benchmarks in table.~\ref{tab:benchmarks}.}
    \label{tab:benchmarks2}
\end{table}

If the the power spectrum of $\chi$ contains features that are not seen in the CMB, $\zeta_{\chi}\f{PT} < 4.5 \times 10^{-5}$ is required (see for example \cite{Bodas:2022zca}). 
In table.~\ref{tab:benchmarks2}, we consider another set of benchmarks in which the PT sector only makes up about $10\%$ of the late-time SM+DM density perturbations with $\zeta_{\chi}\f{PT} = 4.5 \times 10^{-6}$.
The advantage of our eMD model is even more apparent in this case where now the GWB inhomogeneities are larger in the eMD model for all $\zeta_{\chi} \gtrsim 10^{-4}$.
A smaller $\f{PT}$ in this case would require longer period of eMD, and possibly lower $T_{\rm SM}(t_{\rm dec})$ at the end. This can be accomplished by considering an appropriate DM model that freezes-out at lower temperature.

\begin{figure}[ht]
    \centering
    \subfloat[\footnotesize{Isotropic part}]{\includegraphics[width=0.7\columnwidth,trim={3.8cm 5.5cm 6cm 5.cm},clip]{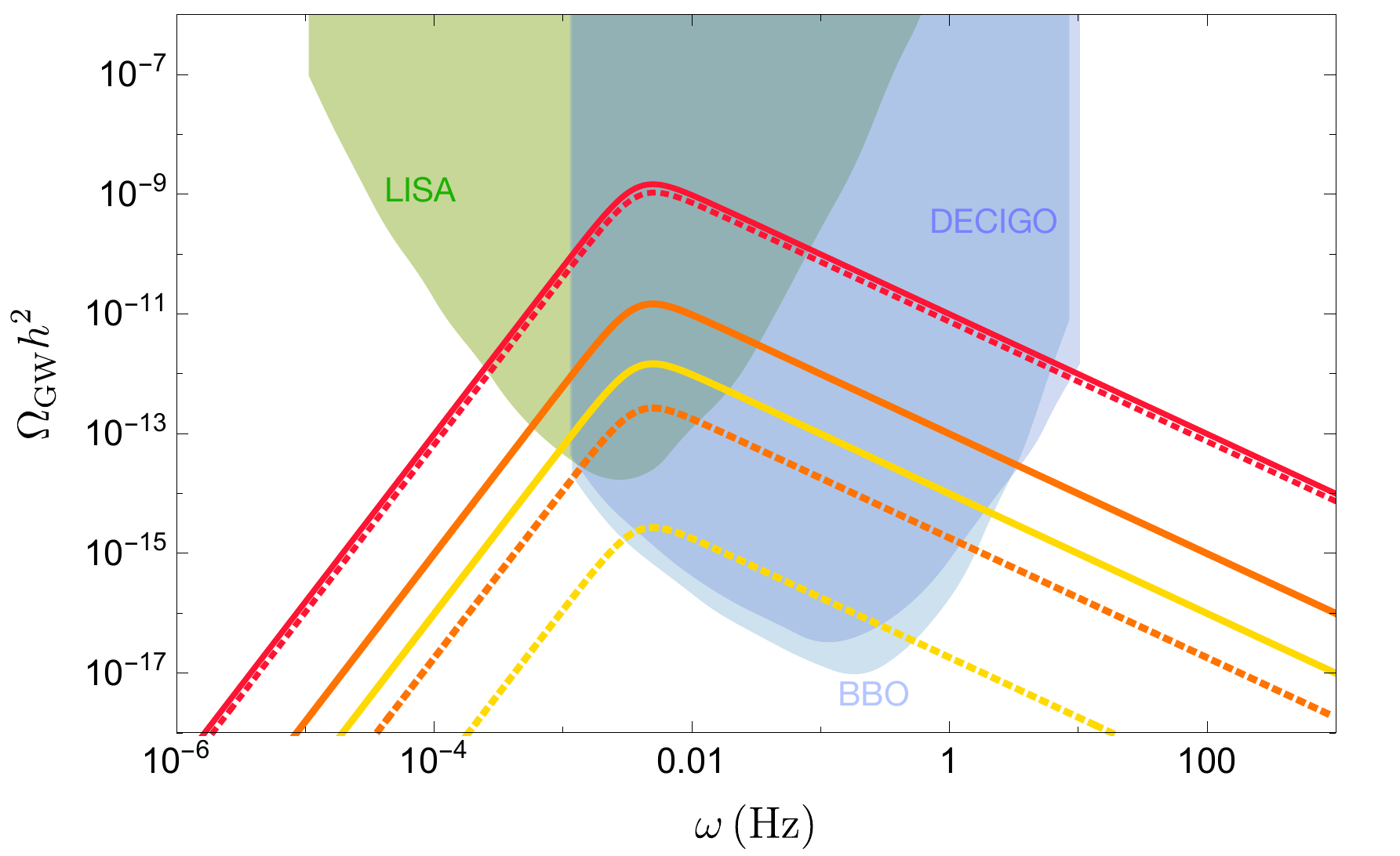}}
    \hspace{1em}
    \subfloat[\footnotesize{Anisotropies}]{\includegraphics[width=0.7\columnwidth,trim={4.5cm 4.3cm 4.9cm 5.3cm},clip]{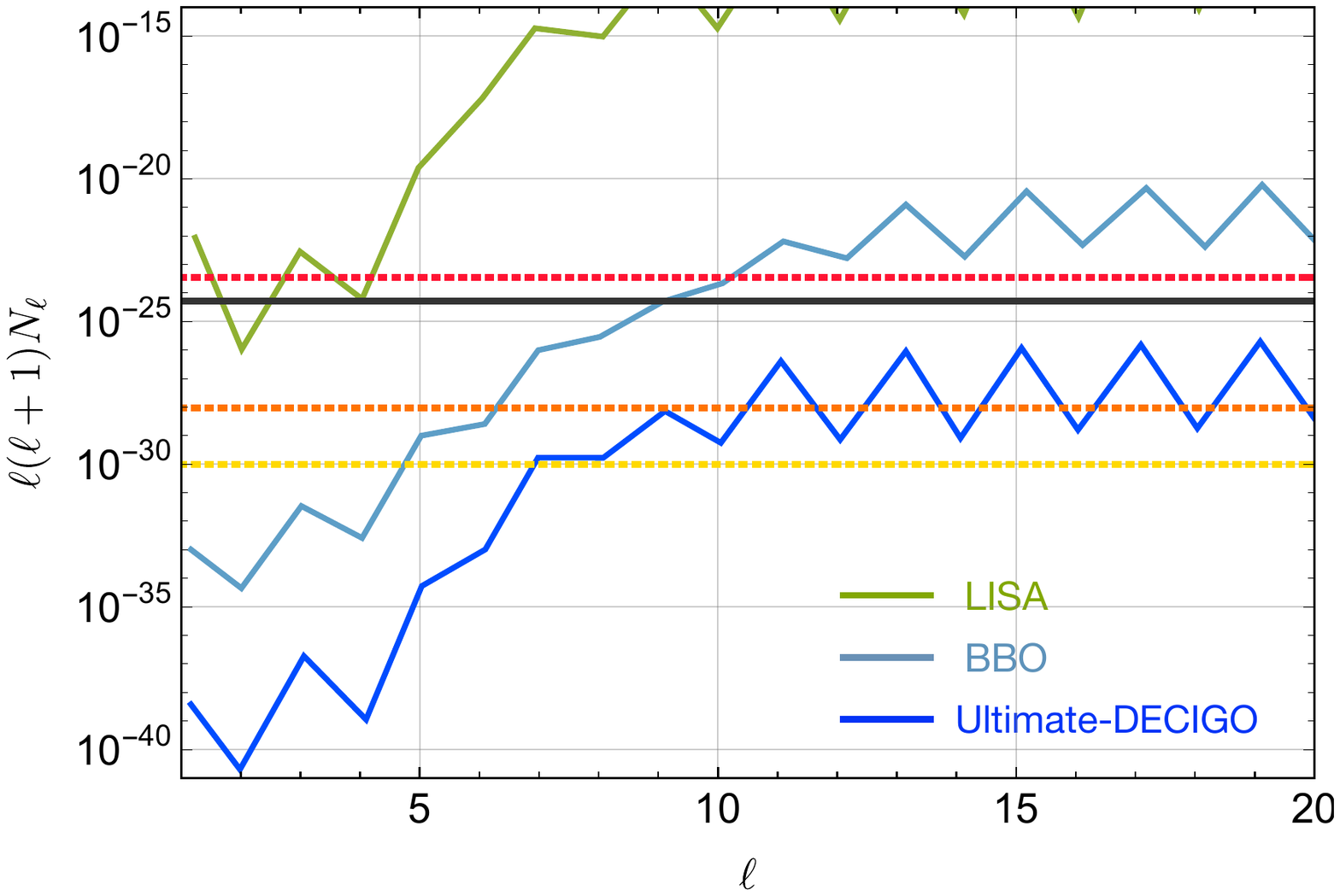}}
	\caption{Comparison of benchmarks 1 (red), 3 (orange), and 4 (yellow) from table.~\ref{tab:benchmarks} with the (a) sensitivities of upcoming space-based experiments like LISA, BBO, and DECIGO, and (b) noise power spectra $N_{\ell}$ as computed in \cite{Braglia:2021fxn}. The solid lines correspond to our modified eMD model while the dotted lines correspond the radiation-dominance scenario of \cite{Geller:2018mwu}. In (b), all benchmarks for the eMD model lie on the same line at $\approx 5\times 10^{-25}$ and are together represented by a black line.} 
    \label{fig:benchmarks}
\end{figure}

To access the improvement in the detection prospect of anisotropies with the eMD model, let us compare the angular power spectrum of GWB to the corresponding angular sensitivity of experiments. 
The anisotropies of PT-GWB can be conveniently expressed in the multipole basis analogous to the CMB,
\begin{align}
    \langle \delta\Omega_{\rm GW}^{(0)}(\hat{n}) \delta\Omega_{\rm GW}^{(0)}(\hat{n}') \rangle
    =  \sum_{\ell} \frac{(2 \ell +1)}{4 \pi} C^{\rm GW}_{\ell} P_{\ell}(\cos \theta)\, , 
\end{align}
where  $\hat{n}\cdot\hat{n}' = \cos \theta$ and $P_{\ell}$ are the Legendre polynomials.
The angular power spectrum $ C_{\ell}^{\rm GW} $ can be related to the amplitude of GW fluctuations as, 
\begin{align}\label{eq:CellIntegral}
    C_{\ell}^{\rm GW} = \frac{2}{\pi} \int_{0}^{\infty} \frac{dk}{k} {\cal P}_{\delta\Omega_{\rm GW}}(k) j^{2}_{\ell}[k (\eta_{\rm PT}-\eta_{\rm today})],
\end{align}
where $j_{\ell}$ are spherical Bessel functions and ${\cal P}_{\delta\Omega_{\rm GW}}$ is given by
\begin{align}
    \langle \delta\Omega_{\rm GW}^{(0)}(\vec{k}) \delta\Omega_{\rm GW}^{(0)}(\vec{k}') \rangle = \delta(\vec{k}+\vec{k}') \frac{(2 \pi)^3}{k^3} {\cal P}_{\delta\Omega_{\rm GW}}(k).
\end{align}
For a scale-invariant spectrum, ${\cal P}_{\delta\Omega_{\rm GW}} (k)$ is constant.
In this case, the integral in \Eq{eq:CellIntegral} can be simplified \cite{dodelson2020modern}, giving
\begin{align}\label{eq:CellSimplified}
    \ell(\ell+1)C_{\ell}^{\rm GW} \approx 14 {\cal P}_{\delta\Omega_{\rm GW}} 
\end{align}
$C_{\ell}^{\rm GW}$ can be compared to the noise power spectrum $N_{\ell}$ of the experiments.
Figure~\ref{fig:benchmarks}~(b) shows $\ell(\ell+1)N_{\ell}$ as a function of multipole $\ell$ for relevant detectors like LISA, BBO, and ultimate-DECIGO, which where computed in ref.~\cite{Braglia:2021fxn}, and are based on the code of ref.~\cite{Alonso:2020rar}. The detector design gives higher sensitivity to even multipoles, resulting in the zigzag curves in the plot.   

Fig.~\ref{fig:benchmarks} shows the strength of GWB signals for benchmarks 1 (red), 3 (orange), and 4 (yellow) from table~\ref{tab:benchmarks} in the radiation-dominated (dotted lines) and eMD model (solid lines).
Fig.~\ref{fig:benchmarks}~(a) compares the strength of the GWB monopoles with the projected power-law integrated sensitivities for LISA, BBO, and DECIGO \cite{Schmitz:2020syl}.
Fig.~\ref{fig:benchmarks}~(b), on the other hand, illustrates the strength of the anisotropic GWB power spectrum (computed using \Eq{eq:CellSimplified}) in the radiation-dominated and eMD models, against projections of the noise power spectrum for LISA, BBO, and ultimate-DECIGO configurations \cite{Braglia:2021fxn}. 
As explained before, there is no additional $\f{PT}$ suppression in the GWB inhomogeneities in the eMD model, and the power spectrum for all the benchmarks lies at the same position $\sim 5 \times 10^{-25}$ shown as a black line in \Fig{fig:benchmarks}~(b).

\section{Discussion}\label{sec:discussion}
LISA is the most sensitive GW detector already under development, with a baseline sensitivity of $\Omega_{\rm LISA} h^2 \sim 10^{-12}$ \cite{amaro2017laser,caprini2016science} and a power-law integrated sensitivity of $\Omega_{\rm LISA}^{\rm int} h^2 \sim 10^{-14}$ \cite{Schmitz:2020syl}. 
Clearly, it would be important to be able to see at least an isotropic GWB signal at LISA in order to motivate
development of more futuristic proposed detectors, such as BBO or DECIGO, to measure the anisotropies. We see from 
\Fig{fig:benchmarks}~(a) and table~\ref{tab:benchmarks2} that for larger fractional GWB anisotropies, even the isotropic component can be challenging at LISA for the radiation-domination models, whereas they can be readily detectable in our early-matter-domination (eMD) models. 

Once an isotropic GWB is detected, there would be a {\it guaranteed} anisotropic component to be discovered. In fact we see in \Fig{fig:benchmarks}~(b) that with our eMD model, it is possible that the first few GWB multipoles might be visible already at LISA, above the instrumental noise anisotropy, for a much larger range of GWB anisotropies.
Detection of a highly anisotropic GWB at LISA would make an even stronger case for more sensitive detectors to mine the physics of the isocurvature GWB map with higher multipoles.
Figure.~\ref{fig:benchmarks}~(b) shows that such high resolution map-making would be possible within our eMD model.
Finally, larger GWB signals in the eMD model will be significantly easier to distinguish from the various astrophysical foregrounds \cite{Cusin:2019jhg}.

\section*{Acknowledgements} 
The authors would like to thank Peizhi Du, Anson Hook, Mrunal Korwar, Soubhik Kumar, Peter Shawhan, Gustavo Marques Tavares, and Lian-Tao Wang  for useful discussions and comments. This work was supported by NSF grant PHY-2210361 and the Maryland Center for Fundamental Physics.

   \bibliographystyle{JHEP}
    \bibliography{refs.bib}

\end{document}